\documentclass[fleqn,usenatbib]{mnras}

\usepackage{newtxtext,newtxmath}
\usepackage{epsfig}

\usepackage[T1]{fontenc}
\usepackage{ae,aecompl}

\usepackage{graphicx}	
\usepackage{amsmath}	
\usepackage{grffile} 

\newcommand{\Ms}{{\ensuremath{\mathrm{M}_{\odot} }}}

\title[IMBHs in the First Galaxies]{Radiation Hydrodynamical Simulations of the Birth of Intermediate-Mass Black Holes in the First Galaxies}

\author[Latif et al.]{Muhammad A. Latif,$^{1}$\thanks{E-mail: latifne@gemail.com} 
Sadegh Khochfar,$^{2}$ 
Dominik Schleicher,$^{3}$ 
Daniel J. Whalen$^{4,5}$\\
$^{1}$Physics Department, College of Science, United Arab Emirates University, PO Box 15551, Al-Ain, UAE \\
$^{2}$Institute for Astronomy, University of Edinburgh, Royal Observatory, Blackford Hill, Edinburgh EH9 3HJ, UK \\
$^{3}$Astronomy Department, Universidad de Concepci\'on, Barrio Universitario, Concepci\'on, Chile \\
$^{4}$Institute of Cosmology and Gravitation, University of Portsmouth, Portsmouth PO1 3FX, UK \\
$^{5}$Ida Pfeiffer Professor, University of Vienna, Department of Astrophysics, Tuerkenschanzstrasse 17, 1180, Vienna, Austria
}

\date{Accepted XXX. Received YYY; in original form ZZZ}

\pubyear{2020}

\begin{document}
\bibliographystyle{mnras}

\label{firstpage}
\pagerange{\pageref{firstpage}--\pageref{lastpage}}
\maketitle

\begin{abstract}

The leading contenders for the seeds of $z > 6$ quasars are direct-collapse black holes (DCBHs) forming in atomically-cooled halos at $z \sim$ 20.  However, the Lyman-Werner (LW) UV background required to form  DCBHs of 10$^5$ \Ms\ are extreme, about 10$^4$ J$_{21}$, and may have been rare in the early universe. Here we investigate the formation of intermediate-mass black holes (IMBHs)  under moderate LW backgrounds of 100 and 500 J$_{21}$ that were much more common at early times. These backgrounds allow halos to grow to a few 10$^6$ - 10$^7$ \Ms\ and virial temperatures of nearly 10$^4$ K before collapsing but do not completely sterilize them of H$_2$. Gas collapse then proceeds via Ly$\alpha$ and rapid H$_2$ cooling at rates that are 10 - 50 times those in normal Pop III star-forming haloes but less than those in purely atomically-cooled haloes.  Pop III stars accreting at such rates become blue and hot, and we find that their ionizing UV radiation limits their final masses to 1800 - 2800 \Ms\, at which they later collapse to IMBHs. Moderate LW backgrounds thus produced IMBHs in far greater numbers than DCBHs in the early universe.

\end{abstract}




\section{Introduction}

The discovery of quasars at $z > 7$ \citep{Mortlock11,Wu15,Banados18,Yang20} poses significant challenges to current paradigms of structure formation because it is not fully understood how supermassive black holes (SMBHs) formed less than a Gyr after the Big Bang \citep{Volonteri10,Haiman13,Latif16PASA,Woods19,Inayoshi19}. A number of origins have been proposed for high-$z$ quasars such as the BHs of Pop III stars at $z \sim$ 25 \citep[10 - 500 \Ms;][]{Madau2014,Volonteri2015}, runaway stellar collisions in marginally-enriched halos at $z \sim$ 15 - 20 \citep[1000 - 4000 \Ms;][]{Devecchi2009,sak17,Reinoso18a,boek18}, and direct-collapse black holes (DCBHs) in atomically-cooled halos at $z \sim$ 15 - 20 \citep[$\sim$ 10$^5$ \Ms;][]{Schleicher13,Latif13c,Regan14B,Smidt17,Chon18,maio19}.

DCBHs form when primordial halos grow to masses $\gtrsim$ 10$^7$ \Ms\ and virial temperatures of $\sim$ 10$^4$ K without having previously formed a star because they are immersed in strong Lyman-Werner (LW) UV backgrounds that destroy their H$_2$ \citep[e.g.,][]{Latif14UV,Agarwal15,Agarwal19} or in supersonic baryon streaming flows that prevent star formation even if H$_2$ is present \citep{th10,Stacy11,Greif11,srg17}. Temperatures of 10$^4$ K activate Ly$\alpha$ cooling that triggers rapid baryon collapse at up to 1 \Ms\ yr$^{-1}$ \citep{Wise2008,Regan09}. Stellar evolution calculations have shown that such flows can create stars with masses $\gtrsim$ 10$^5$ \Ms\ that collapse to DCBHs via the general relativistic instability \citep{Hosokawa13,um16,tyr17,hle17,hle18}. DCBHs are the leading candidates for the seeds of the first quasars because they form in dense environments in halos that can retain their fuel supply even when it is heated by the BH \citep{Whalen04,Alvarez09,Whalen12,Johnson2013b,Smith18,zhu20}.

The accretion rates of 0.1 - 1 \Ms\ yr$^{-1}$ required to build up 10$^5$ \Ms\ stars can only be sustained by atomic cooling, not H$_2$, and the LW backgrounds required for the complete extinction of H$_2$ in halos are extreme, as much as a few 10$^4$ $\rm J_{21}$ where $\rm J_{21}$ $=$ 10$^{-21}$ erg s$^{-1}$ cm$^{-2}$ Hz$^{-1}$ sr$^{-1}$ \citep{Sugimura14,Latif15a}. But there is a growing body of work that suggests that massive stars can form even in the presence of minute amounts of H$_2$ shielded deep in the cores of halos exposed to more modest LW fluxes. \cite{SS12} examined the collapse of a 3 $\times$ 10$^7$ \Ms\ halo in a LW background of 100 $\rm J_{21}$ and found that  cooling at its center was still governed by H$_2$ and produced a sink particle with a mass of $\sim$ 1100 \Ms. \citet{Latif14ApJ} also found that 100 - 10,000 \Ms\ primordial stars could form in halos immersed in moderate LW fluxes \citep[see also][]{LatifV15}. 

Until recently, simulations of baryon collapse in atomically-cooled halos ignored radiative feedback from stars at their centers because stellar evolution models predicted that they were likely cool and red, and therefore not strong sources of ionising UV flux capable of halting accretion onto the star (although pressure due to outflowing Ly$\alpha$ cooling radiation may could affect flows deep in the core of the halo; \citealt{aaron17}). The first models to incorporate stellar feedback found it had little effect on the growth of the star on AU scales over times of a few years \citep{luo18,ard18}. \citet{Chon18} and \citet{Regan18b} followed the collapse of halos in LW backgrounds of 100 \& 1000 $\rm J_{21}$ for a few hundred kyr and also found that radiation from stars did not suppress accretion. However, the presence of even small mass fractions of H$_2$ in halos in LW backgrounds of these magnitudes can reduce infall rates by 1 - 2 orders of magnitude, down to 0.005 - 0.03 \Ms\ yr$^{-1}$. Stars growing at the low end of this range have been found to become blue and hot in stellar evolution models, with ionising UV fluxes that could at least partially quench accretion \citep{hle17}. How such feedback governs the final masses of stars in these backgrounds is not yet known.

The original simulations of atomically-cooled halos that proceeded from cosmological intial conditions did not exhibit fragmentation or the formation of dense clumps that could later become multiple massive stars \citep[although see][]{Bromm03}. Later studies at higher resolution revealed that atomically-cooled gas fragmented on AU scales but could only follow the evolution of the clumps for a few tens of years and could not determine if they later became stars or were subsumed by the disc at later times \citep{Becerra14,Becerra18}. \citet{rd18,Regan18b} and \citet{Chon18} studied fragmentation at somewhat lower resolutions out to a few hundred kyr and found that some clumps persisted for long times but could not determine their final fates. Also \cite{rd18} simulated a single rare halo of  $\rm \sim10^7~\Ms\ $ at z=24.7 while we here explore typical halos forming in high z universe. \citet{suaz19} followed the collapse of atomically cooled halos at intermediate resolutions in moderate LW backgrounds for $\sim$ 600 kyr, longer than previous studies but still well short of the collapse of the stars. Inflow rates lasting for the times required to actually form DCBHs have only recently been confirmed to occur in numerical simulations \citep{Latif20,ret20,pat20a}. Although binary and multiple DCBH systems formed in all three studies, which holds important implications for the detection of DCBH mergers by the Laser Interferometer Space Antenna (LISA) in coming decades, they did not include radiation from the stars and may not ultimately be self-consistent.

Here, we investigate the prospect of massive BH seeds formation  in typical primordial halos ranging in mass from 1.5 $\times$ 10$^6$ \Ms\ to 2.3 $\times$ 10$^7$ \Ms\ in LW backgrounds of 100 $\rm J_{21}$ and 500 $\rm J_{21}$, which were much more common in the primordial universe than those required for complete photodissociation of H$_2$. Our simulations are evolved for up to $\sim$ 900 kyr, approximately three times longer than in comparable studies, and include radiative feedback from the stars coupled to hydrodynamics and primordial gas chemistry. We describe our simulations in Section 2, present our results in Section 3, and conclude in Section 4.

\section{Numerical Method}

Our simulations were performed with the Enzo adaptive mesh refinement (AMR) cosmology code \citep{Enzo14}. We initialize our runs with cosmological initial conditions generated by MUSIC \citep{Hahn11} at $z$=150 with cosmological parameters taken from Planck 2016 data: $\Omega_{\mathrm{M}}=$ 0.308, $\Omega_{\Lambda}=$ 0.691, $\Omega_{\mathrm{b}} = $ 0.0223 and $h =$ 0.677 \citep{Planck2016}. Our simulation volume is 1 cMpc $h^{-1}$ on a side with a top grid resolution of $256^3$ and two additional nested grids each with a resolution of $\rm 256^3$ that span 20\% of the top grid. We place the halo of interest at the center of the box and allow up to 16 additional levels of refinement during the runs to achieve resolutions of up to $\sim$ 300 AU. We split dark matter (DM) particles in this region into 13 child particles, which produces an effective DM resolution of 5 $\rm M_{\odot}/h$. Beginning at $z=$ 150, we refine on Jeans length, baryonic overdensity and particle mass resolution, as in \citet{Latif20b} and \citet{Latif20}. The Jeans length is resolved by at least 32 cells in our models.

\begin{figure*} 
\begin{center}
\centering
\includegraphics[scale=1.0]{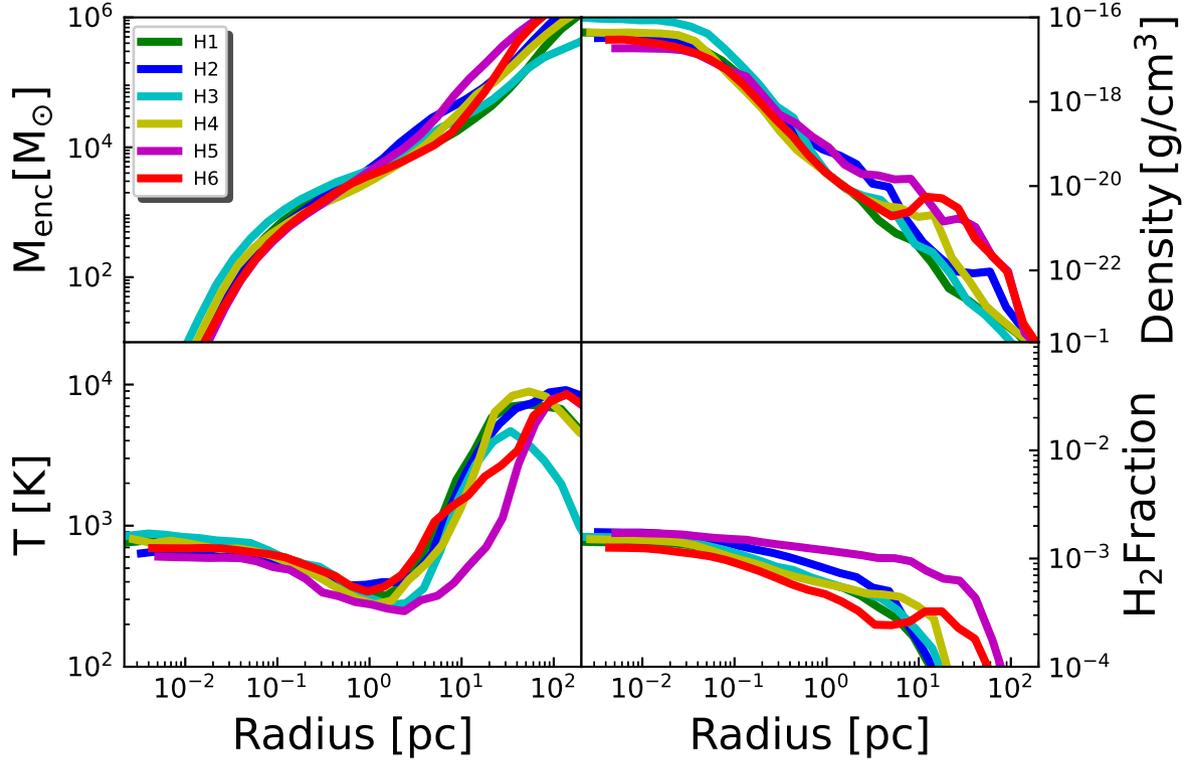}
\end{center} 
\caption{Spherically averaged radial profiles of density, temperature, enclosed mass and $\rm H_2$ mass fraction at the onset of SF. Green, blue, cyan, yellow, magenta and red colors are halos H1, H2, H3, H4, H5, and H6, respectively.}
\label{fig:iprofs}
\end{figure*}

We introduce sink particles that represent Pop III stars in cells where the maximum refinement level is reached, typically at densities of $\rm  \geq 10^{-16}~g/cm^3$.  Our criteria for sink particle formation is based on SmartStars \citep{Regan18b}. A sink particle is allowed to form in a grid cell when meeting the following criteria: {I) the grid cell is at the maximum refinement level; (II) the cell density is higher than the Jeans density; (III) the flow is convergent; (IV) the cooling time is shorter than the free-fall time; (V) the cell is at a local minimum of the gravitational potential. It can accrete gas from a radius of 4 cells and merges with the most massive particle within the accretion radius. Our recipe for accretion is based on the mass influx at the accretion sphere and we use the averaged accretion rate over 1 kyr as the actual accretion rate.

Stars are treated as red and cool if their accretion rates exceed 0.04 \Ms\ yr$^{-1}$ and  blue and hot if they accrete below this rate. We assign a 10$^5$ K blackbody spectrum to blue stars $\rm M_{\odot}$ \citep{Schaerer02} and a 5500 K blackbody spectrum to red stars \citep{Hosokawa13}, assuming in both cases that their luminosities scale with their masses. Consequently as stars grow in mass they become more luminous and produce strong feedback. Photons from stars are propagated throughout the simulation volume with the MORAY raytracing radiation transport package, which is self-consistently coupled to hydrodynamics and non-equilibrium primordial gas chemistry in Enzo \citep{Wise11}. 
Each star is a point source of both ionising and dissociating radiation whose spectrum  is partitioned into five energy bins: 2.0 eV and 12.8 eV, which can destroy H$^-$, H$_2$ and H$_2^+$, and three ionising energies, 14.0 eV, 25.0 eV, 200.0 eV.  We use SEDOP to populate the last three bins which allows us to compute the optimum number of energy bins required to  model radiation above the hydrogen  ionisation  limit \citep{Mirocha12}.  The energy fractions of 0.3261, 0.1073, 0.3686, 0.1965, 0.0 are used in bins 1-5, respectively which are determined from table 4 of \cite{Schaerer02}.

We employ a non-equilibrium primordial chemistry solver \citep{Abel97} to evolve the nine primordial species in our runs: $ \rm H, ~H^+,~ H^-, ~He,~ He^+, ~He^{++},~ H_2, ~H_2^+, ~e^-$. Our simulations include $\rm H_2$ cooling, collisional excitation and ionisation cooling by H and He, bremsstrahlung cooling and recombinational cooling.  Uniform LW backgrounds of 100 or 500 $\rm J_{21}$ are turned on at $z =$ 30 to suppress star formation in minihalos and self-shielding of H$_2$ is approximated by fits from \citet{WG11}.  Such flux is higher than the expected background at  this redshift but could be provided by  star forming galaxies  in the vicinity of the halo. In such scenarios the LW flux becomes almost uniform in the target halo, also see \citet{Agar17}. We ignore radiation pressure due to photoionisations, which could facilitate H II region breakout by imparting momentum to the surrounding gas. \cite{Whalen08} examined the impact of radiation pressure at lower densities and found that they amount to at most 20\% of the ionised gas pressure.  In fact, we performed a test simulation by including radiation pressure and and found the stellar masses to be similar to the case without radiation pressure. We therefore conclude that radiation pressure will not strongly affect our results. 

\begin{table}
\begin{center}
\caption{Virial masses, collapse redshifts and LW backgrounds for the six halos in our study. The final masses of the stars in each halo are listed in column 5.}
\begin{tabular}{| c | c | c | c| c|}
\hline
\hline
Halo & $z$  & Mass (\Ms) & $\rm J_{21}$ & Stars (\Ms) \\
\hline                                                          
H1 &  16.5    &  $\rm 7.5 \times 10^6$  &  100   &  196, 97, 27,  2775       \\
H2 &  14.5    &  $\rm 1.7 \times 10^7$  &  500   &  1668,10, 8683             \\
H3 &  22.8    &  $\rm 1.5 \times 10^6$  &  100   &  2743                            \\ 
H4 &  18.5    &  $\rm 5.4 \times 10^6$  &  500   &  60, 94 ,25, 2638          \\ 
H5 &  13.3    &  $\rm 1.3 \times 10^7$  &  100   &  346, 520, 1161, 2079  \\ 
H6 &  13.31  &  $\rm 1.7 \times 10^7$  &  500   &  546,1955                     \\
\hline
\end{tabular}
\label{tab:tbl-1}
\end{center}
\end{table}

\section{Results}

\subsection{Halo properties}
We have simulated six halos, labelled H1, H2, H3, H4, H5 and H6, whose masses and collapse redshifts are listed in Table~\ref{tab:tbl-1}. Our  halos form in different environments with DM overdensities of 10-100 times the cosmic mean and also have a variety of merger histories.  We compute  the DM overdensity by calculating the mean DM  density in a radius that is 10 times the virial radius of the halo and compare it to the cosmic mean in a volume of $\rm 1~(Mpc/h)^3$. This allows us to estimate the environment of the halo.  H1-H6 have DM overdensities of 10, 50, 90, 70, 100, 65, respectively. Halos H2, H4 \& H5 have undergone a major merger while the others have grown mainly through accretion from the cosmic web and minor mergers. The merger ratios are for H2, H4 \& H5 are 1:3, 1:5 and 1:2, respectively. The halo spins for H1-H6  are 0.02, 0.016, 0.029, 0.012, 0.038, 0.032, respectively.  Their spin parameters vary from 0.015-0.04, which is close to the mean value of 0.02 expected for the halo spin distribution at z $\sim$ 15 \citep{Davis09}. 

\begin{figure*}
\includegraphics[trim=0cm 0cm 0cm 0cm,clip=true,scale=0.662]{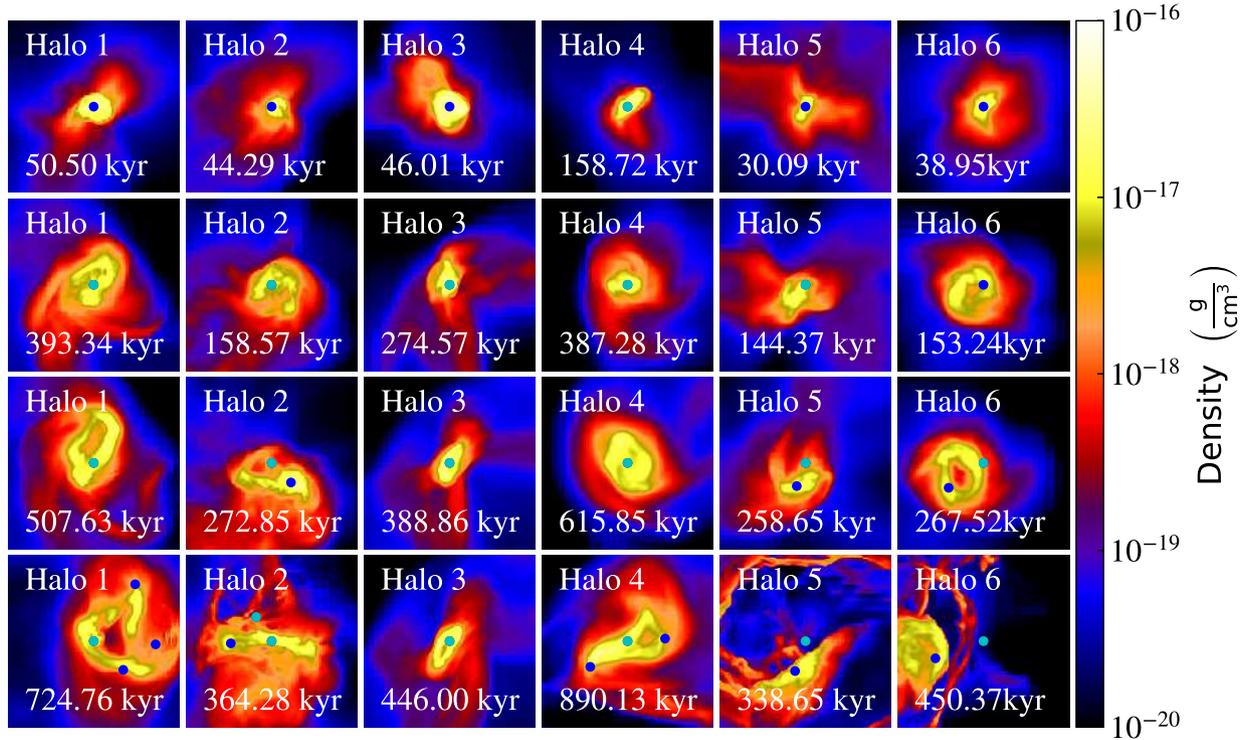}
\caption{Disc evolution and SF in all six halos. The images are density projections showing the average gas density along the line of sight and are 1 pc on a side.  Dark blue dots are stars with masses $\geq$ 10 \Ms\ and cyan dots are stars with masses $\geq$ 1000 \Ms. Each column shows the disc evolution in one halo.}
\label{fig:rho}
\end{figure*}

\subsection{Initial Collapse}

Baryon collapse is suppressed by LW radiation in all six halos until they reach masses of at least 10$^6$ \Ms, and in most cases 10$^7$ \Ms\ \citep{Latif19}. However, as shown in the temperature profiles in Figure~\ref{fig:iprofs}, once collapse begins it is mediated by both Ly$\alpha$ and H$_2$ cooling. At the onset of star formation (SF) atomic cooling dominates down to radii of a few pc, where the gas is at temperatures of $\sim$ 8000 K  due to the dissociation by external UV radiation  shown in the Fig. \ref{fig:tgas}, but H$_2$ cooling, which produces temperatures of 300 - 1000 K, dominates at radii less than 1 pc. This can also be seen in the H$_2$ mass fractions, which can exceed 10$^{-3}$ below 1 pc because they are self-shielded from the external background. The H$_2$-cooled core of the halo is at higher temperatures than those in normal Pop III star-forming halos, which are typically 200 - 300 K, because they are subject to higher mass loading from atomic cooling in the surrounding gas. Overall,  H5 has the higher H$_2$ mass fraction and lower temperature than all halos because of the larger cooling rate. In comparison with H5, H6 has a lower H$_2$  fraction and  higher temperature.

The two-phase temperature structure in our halos is due to the LW background, which allows them to grow to larger masses and higher virial temperatures before collapsing but does not completely sterilize them of H$_2$. Because their cores are at higher temperatures than those of normal Pop III star-forming halos, H$_2$ cooling and formation rates, which peak at 1000 - 2000 K, are much higher there \citep{on07}. This leads to central accretion rates that are much higher than those in normal minihalos but less than those in isothermal atomically-cooled halos in much higher LW backgrounds.

\subsection{Disc Evolution / Star Formation}

\begin{figure*} 
\includegraphics[trim=0cm 0cm 0cm 0cm, clip=true,scale=0.662]{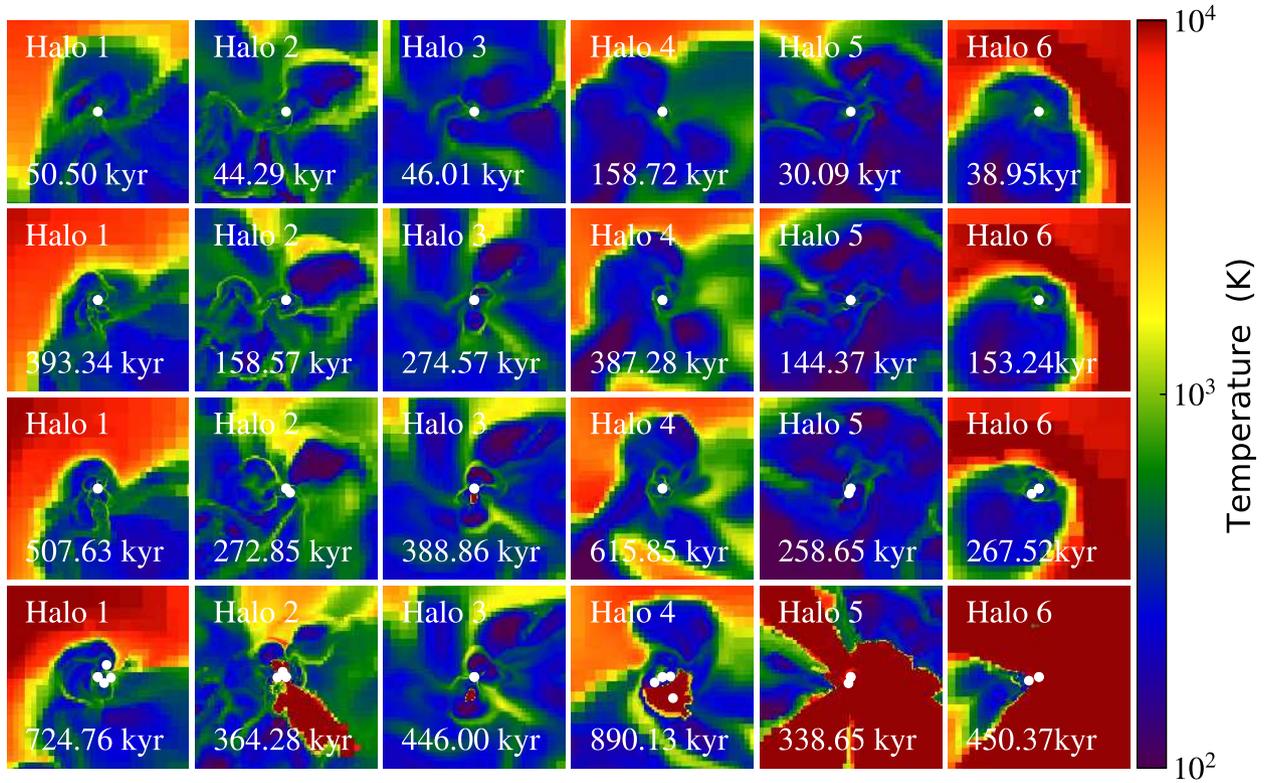}
\caption{Temperature slices of the central 5 pc of the halo, taken along the x-axis.}
\label{fig:tgas}
\end{figure*}

\begin{figure*} 
\includegraphics[trim=0cm 0cm 0cm 0cm, clip=true,scale=0.662]{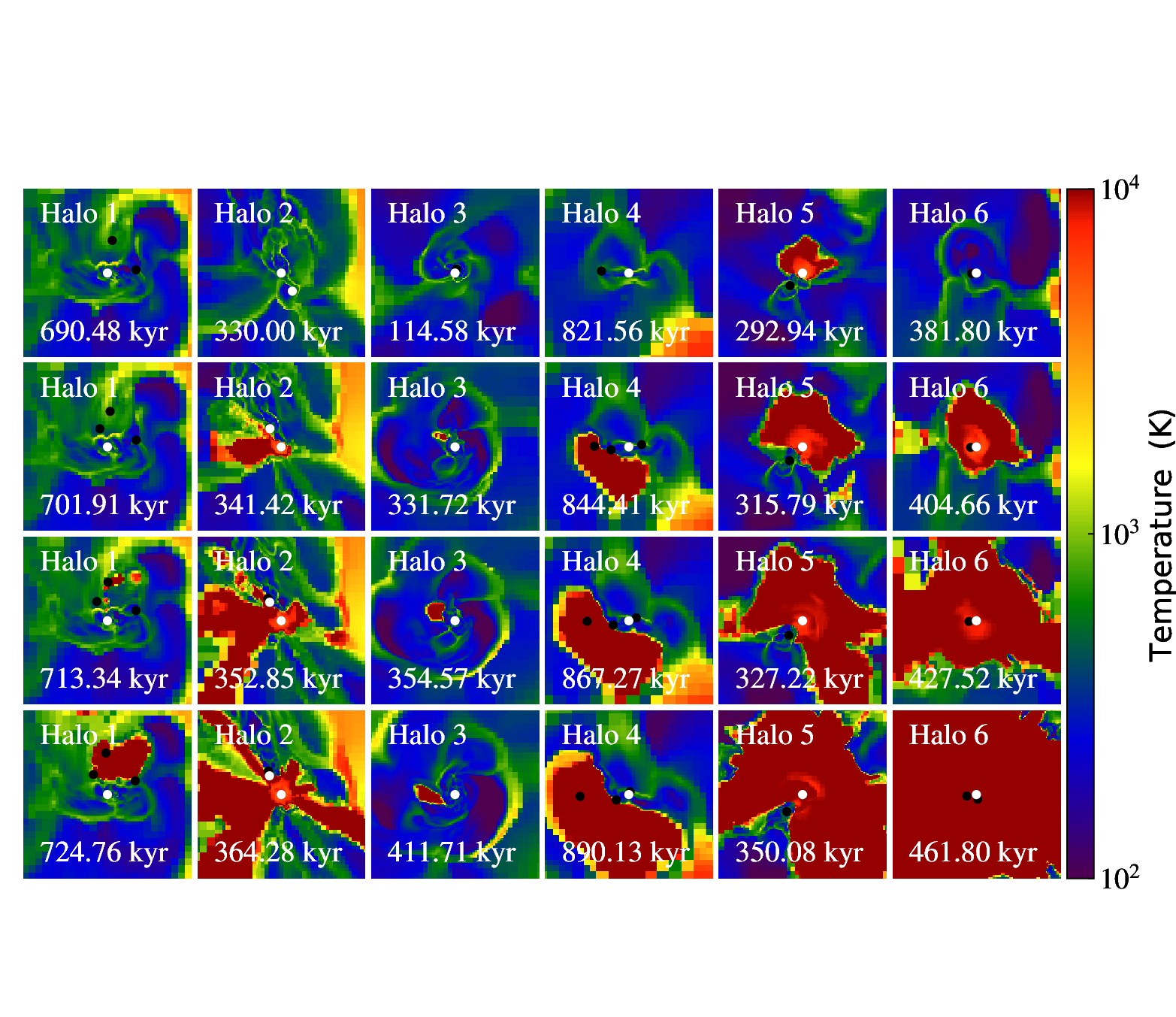}
\caption{The temperature around massive stars is shown  in the central 2 pc of the halo by taking slices along y-axis. In each column we show epochs at the onset and breakout of HII regions  corresponding to  the sharp drops in the mass accretion rates onto stars (< 0.001 $\rm \Ms$/yr).  The white dots are stars with masses $\geq$ 1000 \Ms\ and black dots are stars with masses $\geq$ 10 \Ms.}
\label{fig:hot1}
\end{figure*}

\begin{figure*} 
\includegraphics[trim=0cm 0cm 0cm 0cm, clip=true,scale=0.662]{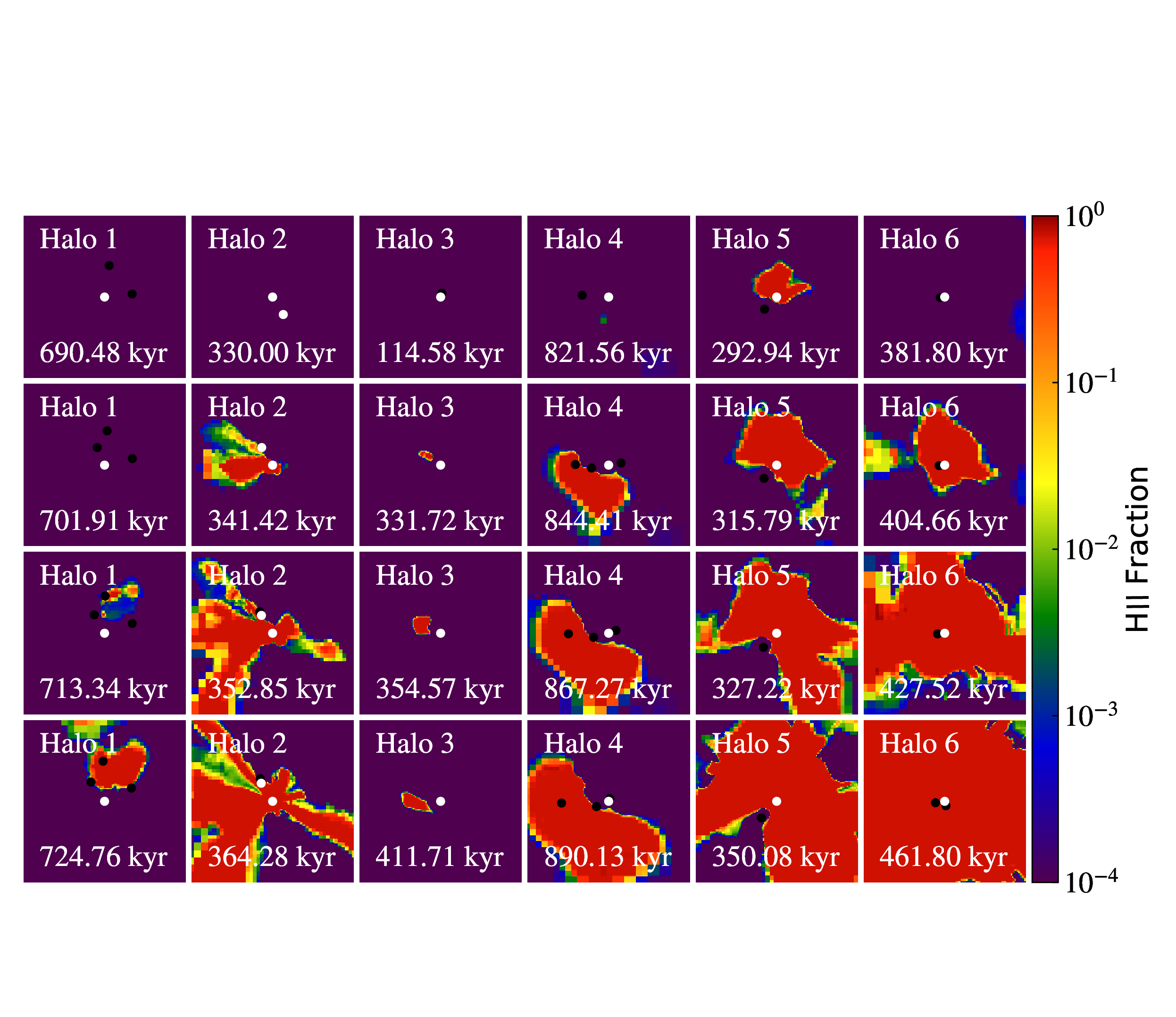}
\caption{The HII fraction around massive stars is shown  in the central 2 pc of the halo by taking slices along y-axis. In each column we show epochs at the onset and breakout of HII regions  corresponding to the  sharp drops in the mass accretion rates onto stars (< 0.001 $\rm \Ms$/yr).The white dots are stars with masses $\geq$ 1000 \Ms\ and black dots are stars with masses $\geq$ 10 \Ms.}
\label{fig:hot2}
\end{figure*}

Discs in our simulations remain stable most of the times with Toomre Q > 1  but occasionally  become unstable and lead to  star formation in Jeans unstable cells at the minima of local gravitational potential. Overall,  we  find that multiple stars form in almost all simulations  but they get merged on short timescales of a few kyr in agreement with previous studies exploring such conditions \citep{SS12,Latif14ApJ}.
As shown in Figure~\ref{fig:rho}, collapse leads to the formation of thick, rotationally flattened discs with initial radii of $\sim$ 0.1 pc and masses of a few thousand solar masses. They grow to $\sim$ 0.2 - 0.3 pc in radius by the end of the runs at nearly 1 Myr. These discs are a factor of five smaller in radius than those in isothermal atomically-cooled flows at similar times because of lower infall rates due to less efficient H$_2$ cooling. At later times the discs develop irregular features created by  ionising UV feedback, as we discuss in greater detail below.

A single star particle with a mass of a few solar masses forms first at the center of each disc when it reaches densities of $\sim$ 10$^{-16}$ g cm$^{-3}$. Although new star particles can form within the accretion radius of this star at times and merge with it a few hundred or thousand years later, it mostly grows by accretion of dense gas. Secondary stars begin to appear by $\sim$ 250 kyr in H2, H5 and H6, at $\sim$ 500 kyr in H1 and at $\sim$ 700 kyr in H4. At the end of the run H1, H4 and H5 host four stars, H2 has three and H6 has two. H3 forms a few low-mass stars but they are subsumed into the central star and only it remains at the end of the run at 900 kyr. Most of the additional stars have masses of a few tens to hundreds of solar masses, as shown in Table~\ref{tab:tbl-1}. 


\subsection{Radiative Feedback}

\begin{figure*}
\begin{center}
\begin{tabular}{cc}
\epsfig{file=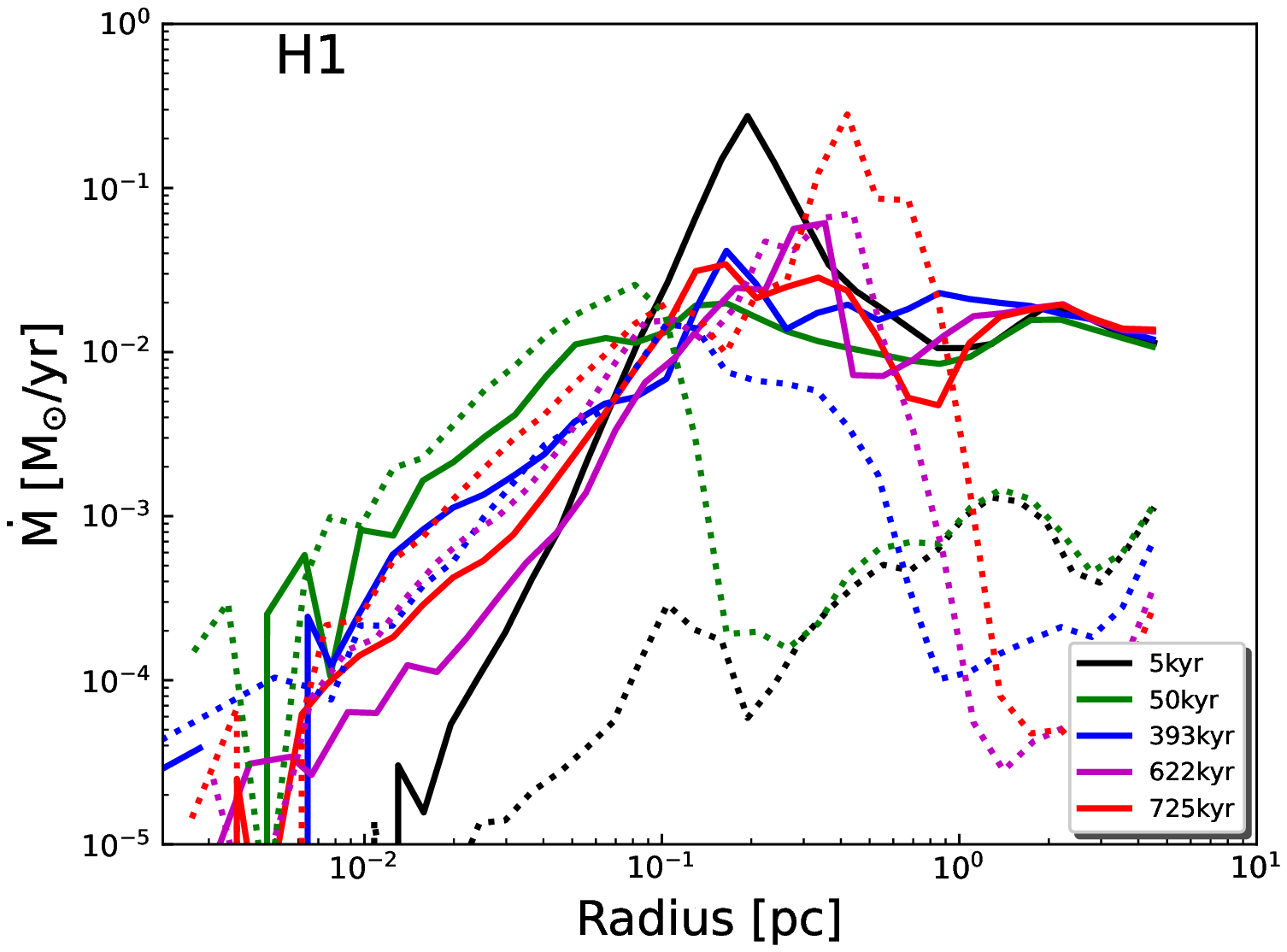,width=0.49\linewidth,clip=}  &  
\epsfig{file=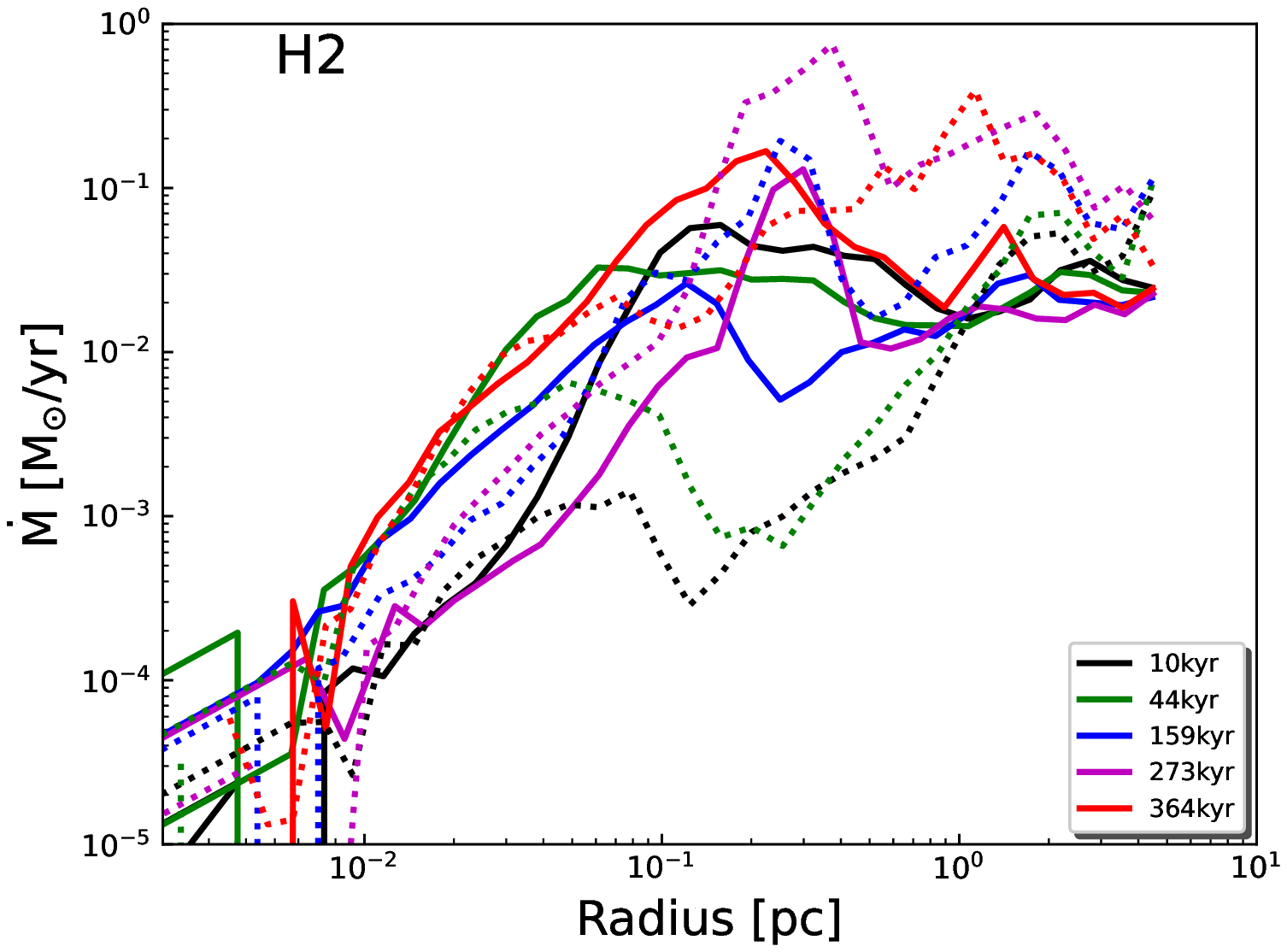,width=0.49\linewidth,clip=}  \\
\epsfig{file=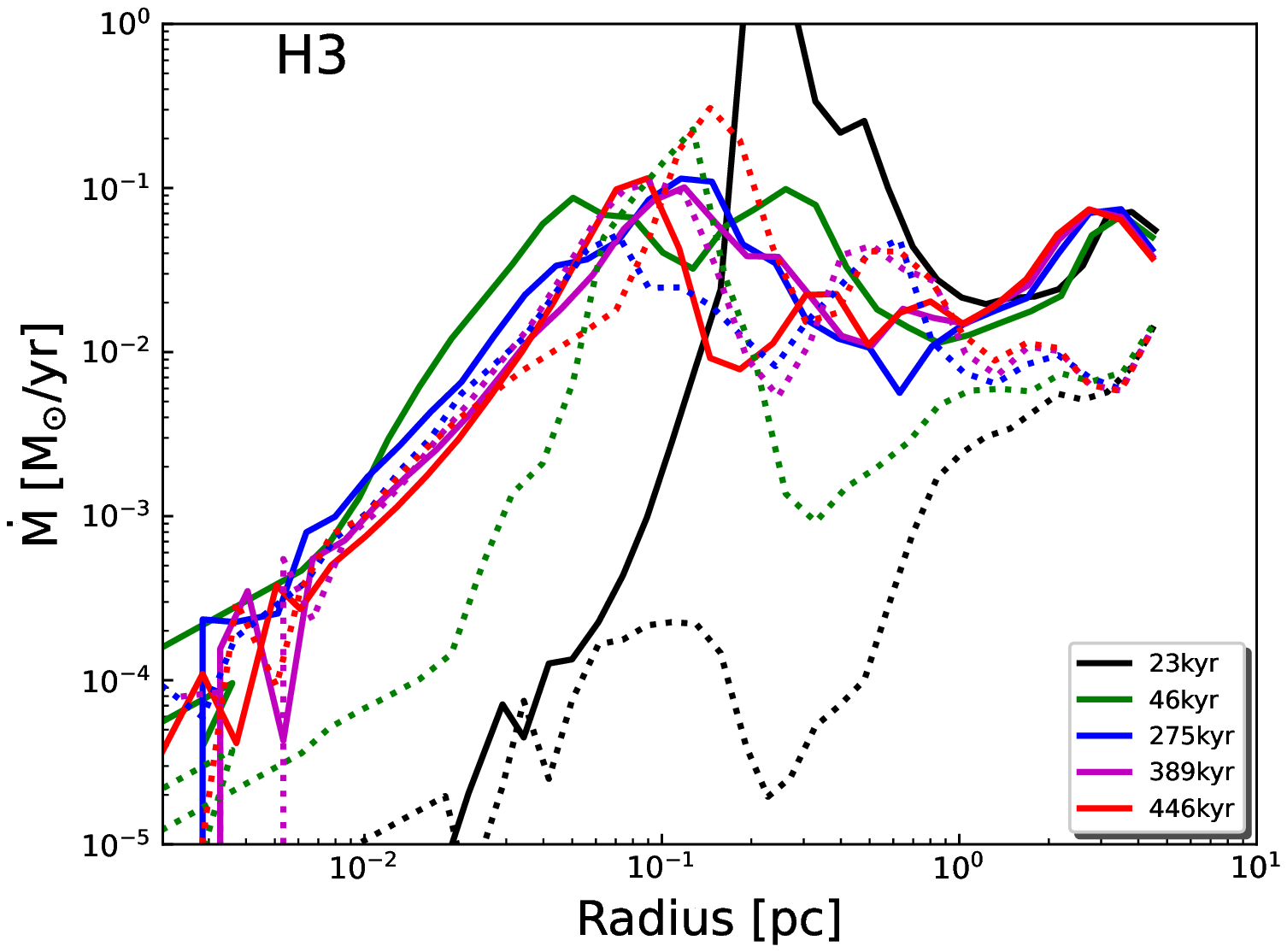,width=0.49\linewidth,clip=}  &  
\epsfig{file=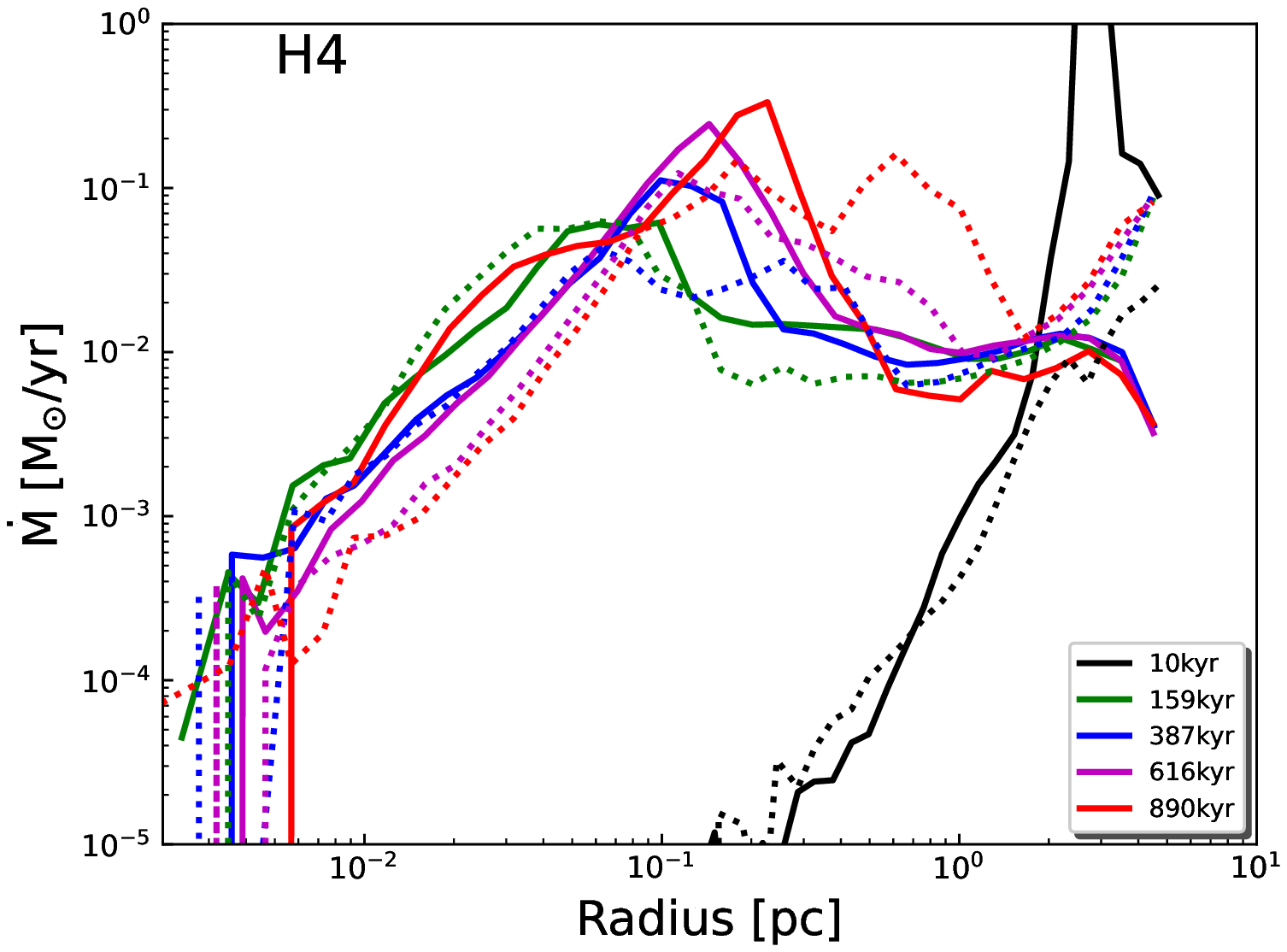,width=0.49\linewidth,clip=}  \\
\epsfig{file=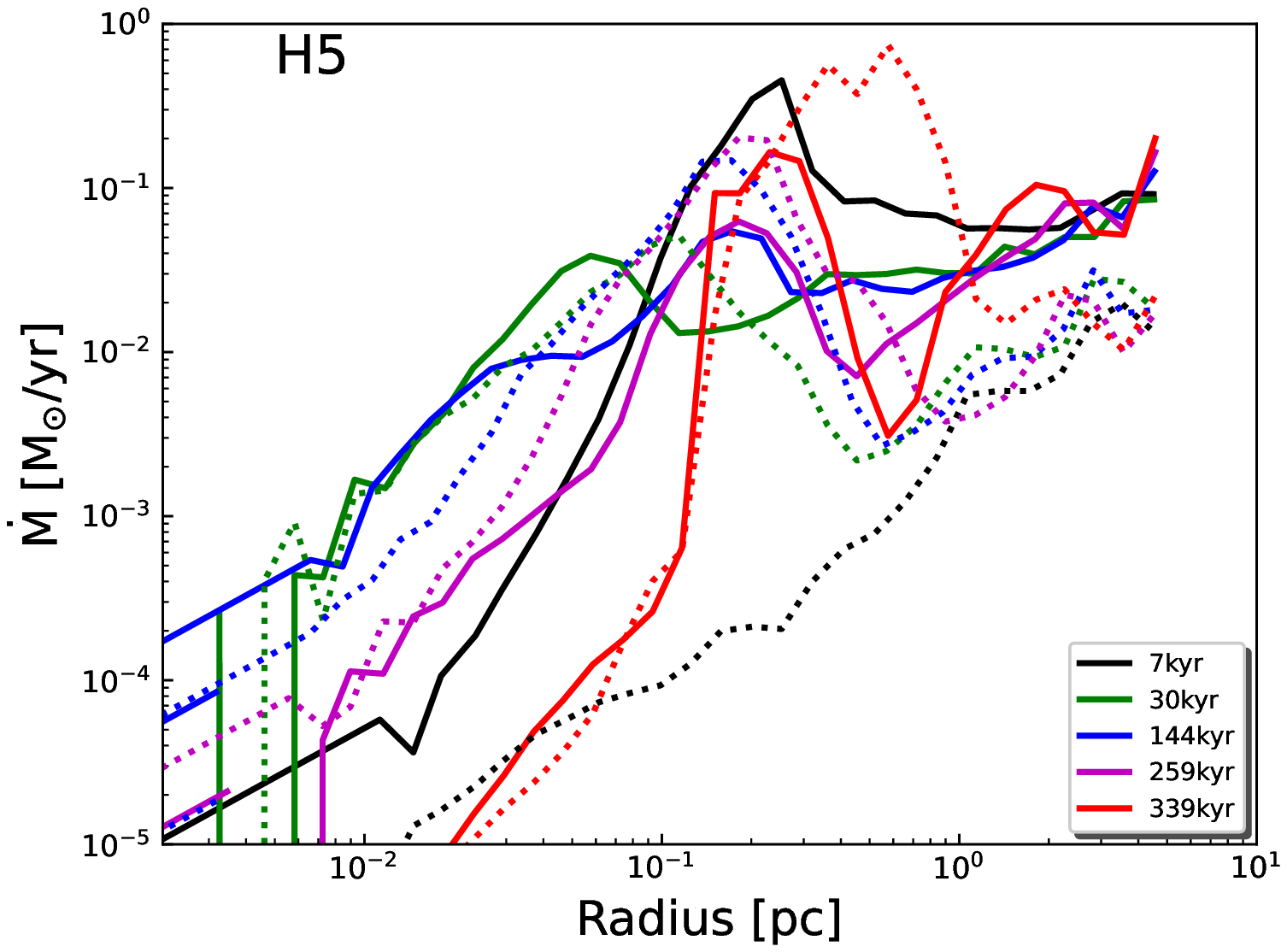,width=0.49\linewidth,clip=}  &  
\epsfig{file=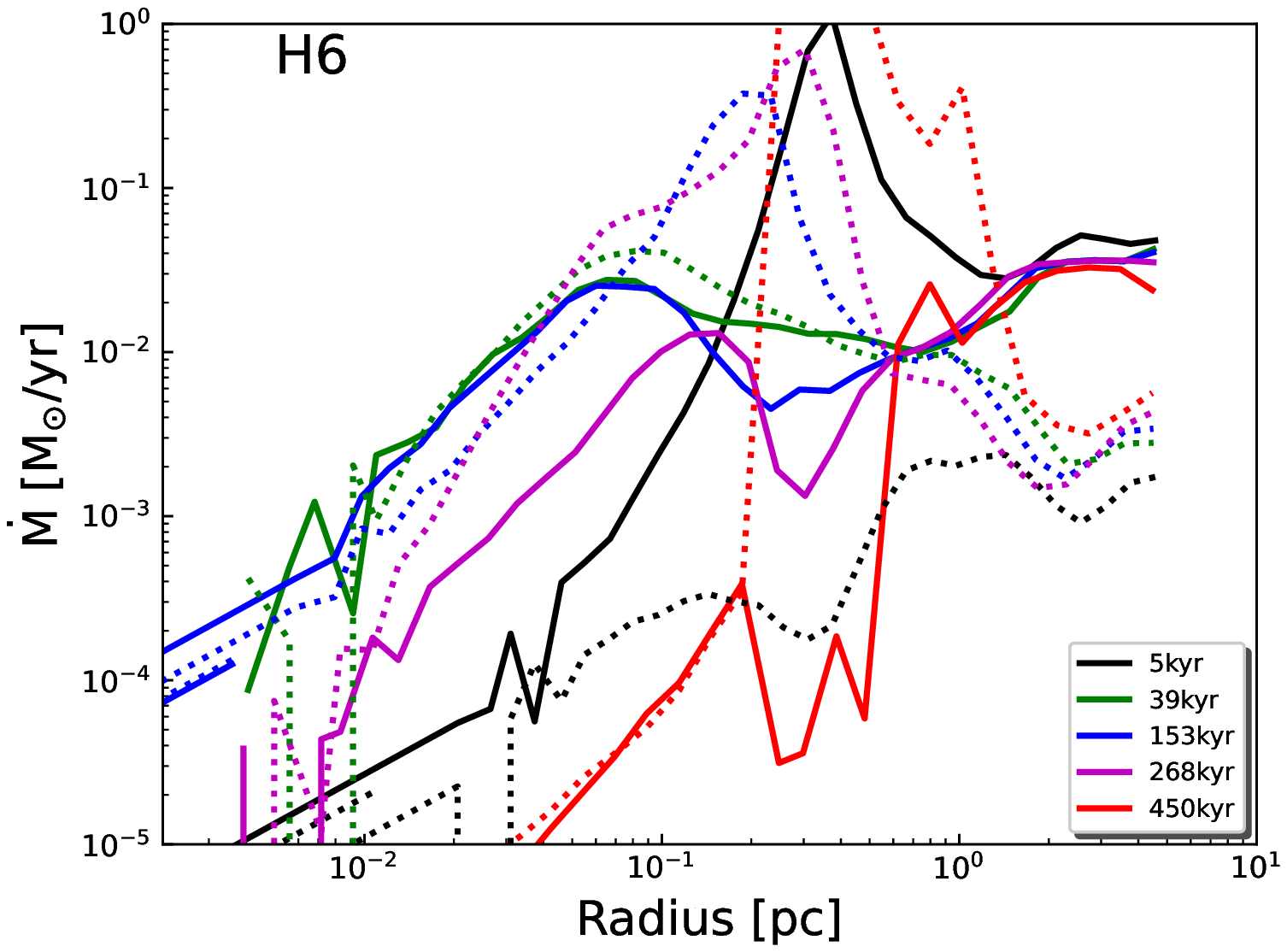,width=0.49\linewidth,clip=}  
\end{tabular}
\end{center}
\caption{Time evolution of spherically averaged gas inflow and outflow rates in radial shells centered on the the most massive star in each halo. Time is calculated  from the formation of the first star in each simulation. Solid lines are inflow rates and dotted lines are outflow rates. Upper left: H1, 100 $\rm J_{21}$; upper right: H2, 500 $\rm J_{21}$; center left: H3, 100 $\rm J_{21}$; center right: H4, 500 $\rm J_{21}$; bottom left: H5, 100 $\rm J_{21}$; bottom right: H6, 500 $\rm J_{21}$.}
\label{fig:outfl}
\end{figure*}

\begin{figure*} 
\begin{center}
\includegraphics[scale=1.0]{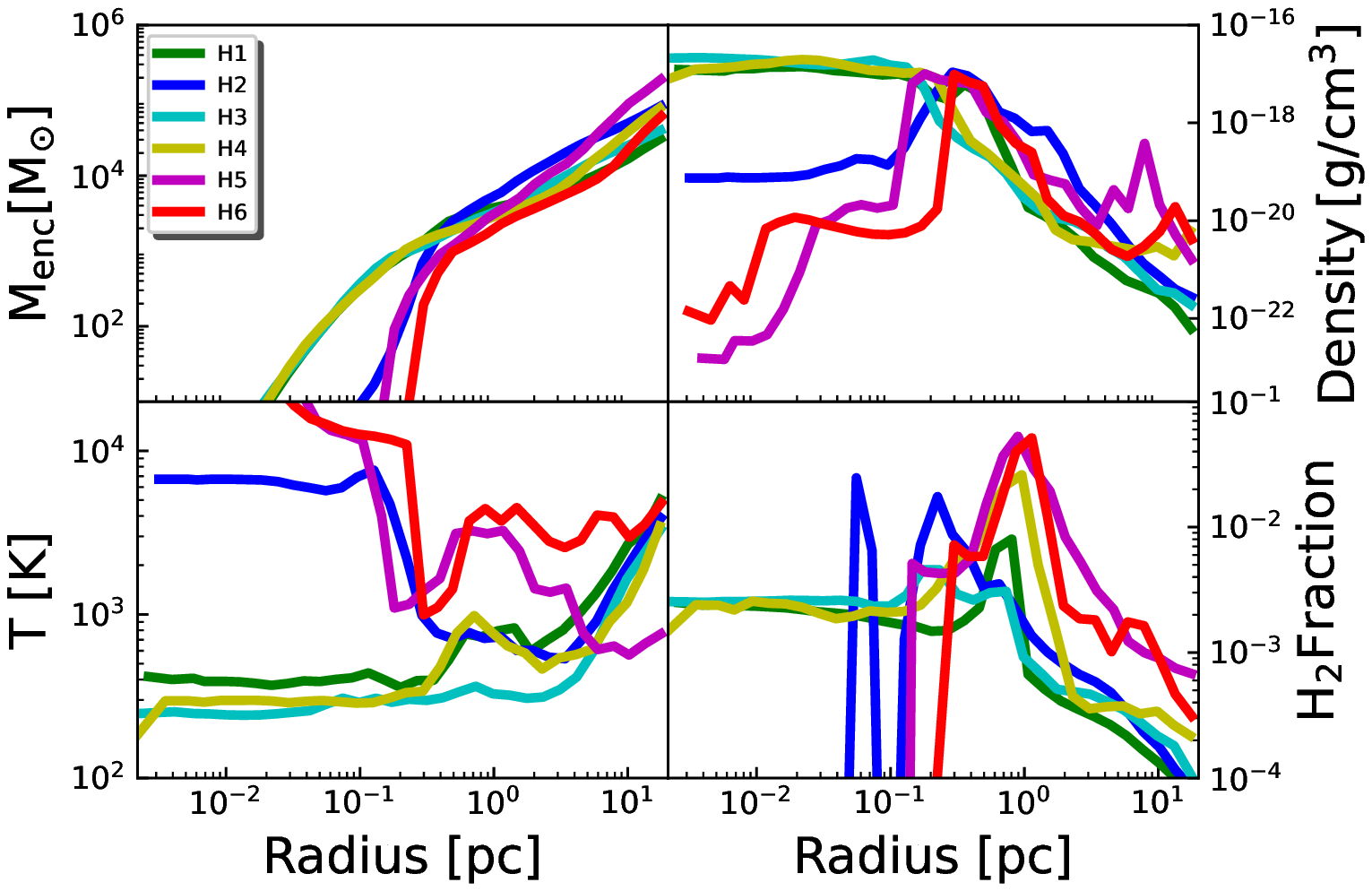}
\end{center} 
\caption{Spherically-averaged densities, temperatures, enclosed masses and $\rm H_2$ mass fractions at the end of each run. Green, blue, cyan, yellow, magenta and red colors are H1, H2, H3, H4, H5, and H6, respectively.}
\label{fig:fprofs}
\end{figure*}

Because accretion onto the main star begins at rates of a few 10$^{-3}$ \Ms\ yr$^{-1}$ in all six halos \citep[on par with normal Pop III star-forming halos;][]{Latif13ApJ,Hosokawa16}, it is initially treated as a hot, blue source of ionising photons. The recombination times in the vicinity of the stars are about 100 yr therefore the UV flux remains trapped deep in the disc. However, the energy deposition mainly by UV ionizing radiation from the central star creates dumbbell  shaped centric outflows which clear the gas from some of the discs, creating the annular structures and cavities in the disks shown in the Fig. \ref{fig:rho}. Stellar feedback lowers the density in the surroundings ($\sim$10$^{-20}$ g cm$^{-3}$) where highly anisotropic H II regions can break out along the lines of sight with lower column densities. The temperatures and ionisation  fractions  in  the HII regions  around the central stars are shown in figures \ref{fig:hot1} and \ref{fig:hot2}, they  have  temperatures of ~ $\rm 2 \times 10^4$ K and HII fractions very close to 1.0. In H2, H5 \& H6 HII regions break out around the central stars at 330 kyrs, 260 kyrs and 400 kyrs, respectively. The I-front blowouts evacuate gas from the vicinity of the  main stars in these three halos and terminate their growth. In H3 the HII region  forms around the primary star at 120 kyrs and  gets detached from the star.  In H1 \& H4 compact HII regions around the central star  get trapped (size of 0.05 pc) and consequently the mass accretion rate onto stars plummets momentarily but  HII regions expand around  secondary stars at 600 kyrs  and 800 kyrs, respectively. In these halos overall energy deposition due to the stellar feedback only declines  mass accretion onto the primary stars by an order of magnitude.  For the central stars,  sizes of the Stromgren radii ($\rm R_S$) vary from 0.6 pc - 3.0 pc due to the anisotropic gas distribution around them and the corresponding Bondi radii ($\rm R_B$) are 1.7 - 2.0 pc. Along the lines of sight with lower column densities  $\rm R_S$ becomes larger than $\rm R_B$ and HII regions break out,  see also \cite{Inayoshi16}.  Conversely, at earlier times $\rm R_S$  < $\rm R_B$ (0.006-0.05 pc vs. 0.09-1.07 pc) for stars with masses of 40 - 500  solar so the H II region cannot grow. Its later expansion is therefore mainly due to the rise in luminosity of the star as it grows by accretion.

To better quantify accretion in the midst of radiative outflows from the discs we plot spherically averaged mass inflow ($\rm 4 \pi R^2 \rho \overline{v_{rad}^-}$) and outflow ($\rm 4 \pi R^2 \rho \overline{v_{rad}^+}$) rates through radial shells in each halo in Figure~\ref{fig:outfl}. Inflow rates average $\sim$ 0.01 \Ms\ yr$^{-1}$ at 1 pc and fall to 10$^{-4}$ \Ms\ yr$^{-1}$ in the vicinity of the main star. Outflows dominate inflows in H2, H5 and H6 due to I-front breakout from the disc. In H3 and H4, outflow rates exceed inflow rates at pc scales at later times. In general, outflow rates are either greater than or comparable to inflow rates within $\sim$ 1 pc of the star in most of the halos. They rise over time, can be quite intermittent, and vary from halo to halo. Inflow rates fall steeply in the  0.2 pc  region around stars due to  stellar feedback,  therefore accretion onto stars is expected to be low at later times.  In the midst of inflows, we estimated the outflow rates shown in Fig \ref{fig:outfl} based on $\rm 4 \pi R^2 \rho \overline{v_{rad}^+}$ but the local expansion of the gas due to turbulence, shocks and even disk dynamics may contribute to outflows, although their contribution is an order of magnitude lower compared to the HII regions.

We show spherically averaged profiles of density, enclosed mass, temperature and $\rm H_2$ mass fraction at the end of our simulations in Figure~\ref{fig:fprofs}. In halos H1, H3 \& H4, the mean densities in the center of the halo have decreased by a factor of a few and the profile has become flattened in the central 0.5 pc  because of the stellar feedback. In H2, H5 and H6 densities have fallen by about six orders of magnitude where shocked ionised flows in the H II region of the star have driven gas out of the core of the halo. Average temperatures in this region have likewise risen above $\rm 10^4$ K because of photoionisations.  $\rm H_2$ mass fractions have fallen by a few orders in magnitude because of photodissociation and collisional dissociation by free electrons. The latter can be important at early stages of H II region formation. The anisotropic H II region of the 1168 \Ms\ star is at an average temperature of 7000 K and has densities and $\rm H_2$ mass fractions that are a few orders of magnitude lower than the surrounding central 0.2 pc of the halo. Average temperatures in the rest of the halos are a few hundred Kelvin and $\rm H_2$ abundances are a few 10$^{-3}$ due to episodic recombination of ionised gas.

The small bumps in the $\rm H_2$ profiles at $\sim$ 0.2 pc and 0.4 pc in H5 and H6 are due to rapid H$_2$ formation in the outer layers of the I-front. The inclusion of multiple ionising photon energies in the spectrum of the star leads to the broadening of the front, and its outer layers can fall to ionisation fractions of 10\% and gas temperatures of 1000 - 2000 K, which are prime conditions for H$_2$ formation in the gas phase via the H$^-$ channel \citep{rgs01,Whalen08}. The larger bumps in H$_2$ abundance at $\sim$ 1 pc in H5 and H6 arise because the dense shell of gas swept up by the D-type I-front behind them partly shields them from LW radiation from the star. Approximately 1000 \Ms\ of gas resides within the central 1 pc in all six halos.

\subsection{Accretion Rates / Final Stellar Masses}

Masses and accretion rates for the main star and next most massive star in each halo are shown in Figure~\ref{fig:mdot}. Accretion rates begin at a few 10$^{-3}$ \Ms\ yr$^{-1}$, rise to a few 10$^{-2}$ \Ms\ yr$^{-1}$, and then fall by an order of magnitude as the stars grow in mass and luminosity. High densities, ram pressures, and short recombination times in the vicinity of the star confine its H II region close to its surface and allow accretion to proceed for a few hundred kyr in all six halos. In H1, an H II region never develops and the star continues to accrete at a few 10$^{-3}$ \Ms\ yr$^{-1}$ until it reaches a mass of 2775 $\rm M_{\odot}$ at the end of the run at 719 kyr. In H2 radiation from the central star terminates its growth 330 kyr after formation at a mass of 1770 $\rm M_{\odot}$. But another star forms at 100 kyr and grows through mergers with other star particles and accretion until it reaches a mass of 8670 $\rm M_{\odot}$ at the end of the simulation at 354 kyr. The two stars produce the highly anisotropic H II region visible at 354 kyr in Figure~\ref{fig:tgas}.

\begin{figure*}
\hspace{-6.0cm}
\centering
\begin{tabular}{c c}
\begin{minipage}{6cm}
\vspace{-0.2cm}
\hspace{-1.2cm}
\includegraphics[scale=0.6]{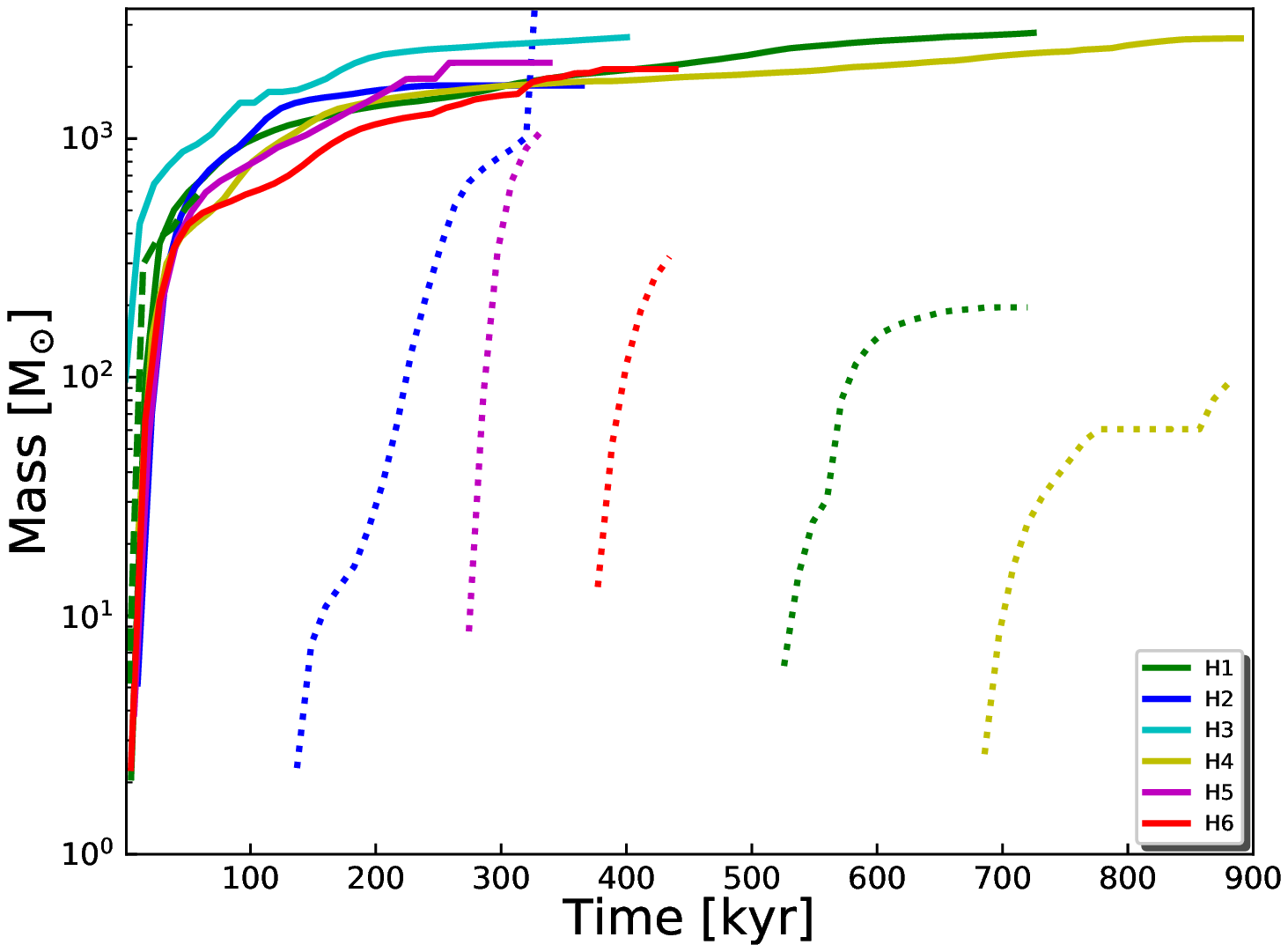}
\end{minipage} &
\begin{minipage}{6cm}
\hspace{1.9cm}
\includegraphics[scale=0.6]{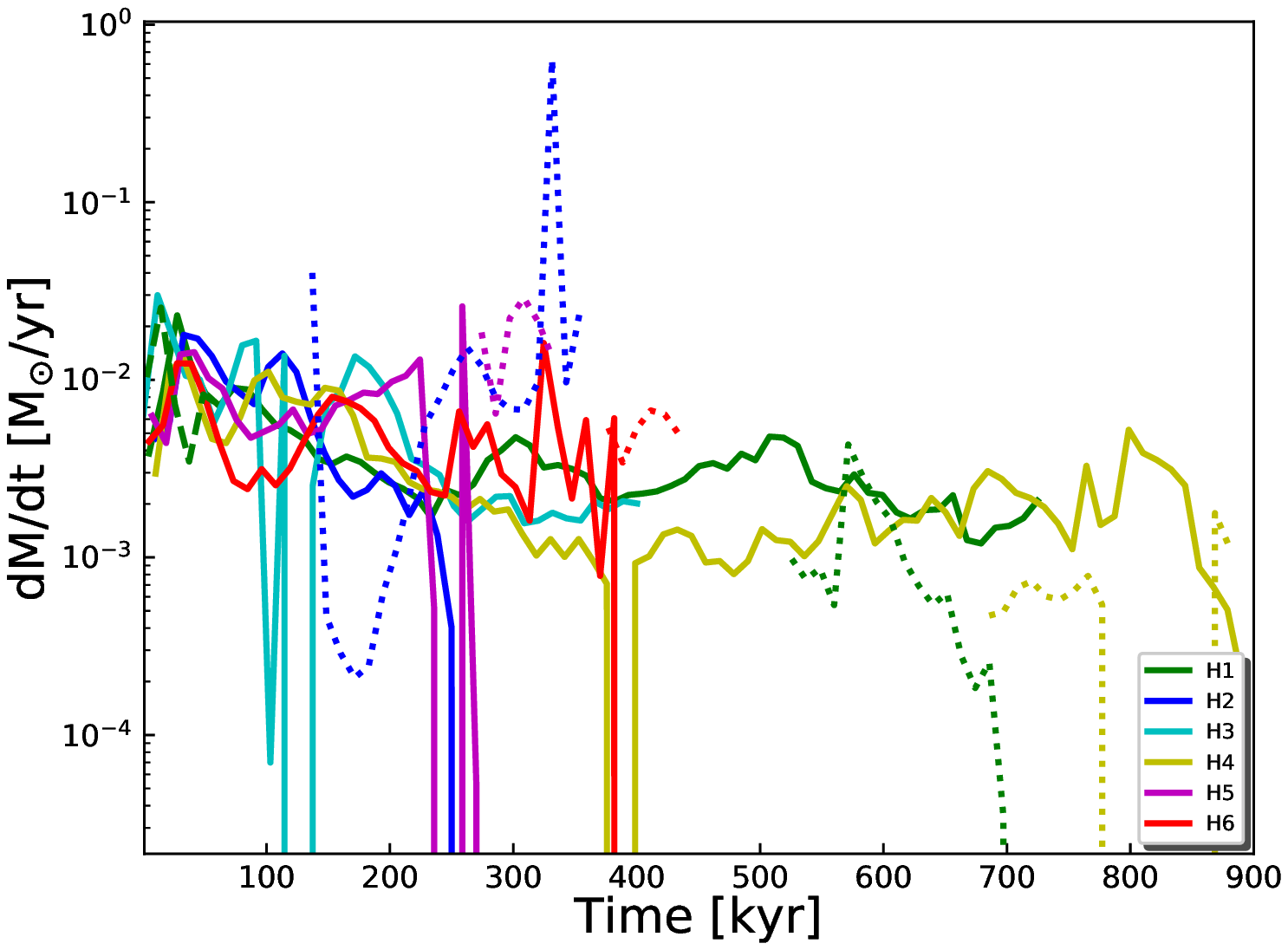}
\end{minipage} 
\end{tabular}
\caption{Accretion rates and masses of the two most massive stars in each halo. The green, blue, cyan, yellow, magenta and red lines are H1, H2, H3, H4, H5, and H6, respectively. The solid lines are for the most massive stars and the dotted lines are for the second most massive star. The dashed green line is for H1 at a resolution of 75 AU evolved to 57 kyr.}
\label{fig:mdot}
\end{figure*}

The star in H3 initially grows faster than those in the other halos, with peak accretion rates of 0.03  \Ms\ yr$^{-1}$ over the first 20 kyr. The large drop in rate at about 100 kyr is due to strong radiatively-driven outflows. The mass of the star at the end of the run is 2743 \Ms, having accreted at an average rate of 0.003  \Ms\ yr$^{-1}$. Average accretion rates for the main star in H4 are similar, $\sim$ 0.003 \Ms\ yr$^{-1}$, and it reaches a mass of 2638 \Ms\ 890 kyr after formation. The sharp dips in rate at 400 kyr and 850 kyr correspond to strong outflows that drive dense gas away from the star, the latter of which is visible in the temperature image at 880 kyr in Figure~\ref{fig:tgas}. In H5 and H6, the initial growth of the stars is comparable to those in the other halos but at 270 kyr and 400 kyr the H II region breaks out of the disc and accretion onto the stars plummets. The final mass of the central star in H5 and H6 is 2079 \Ms\ and 1955 \Ms, respectively.

Overall, accretion onto the stars is intermittent and falls by a few orders in magnitude during outflows. The most massive stars are 1700 - 2800 \Ms\ and, given the flattening in all the mass profiles, are unlikely to grow beyond a few thousand solar masses. This is true of the stars in H1 and H4, which were evolved for twice as long as the others (719 kyr and 880 kyrs, respectively) but grew in mass by less than a factor of two after their profiles flattened out. The second most massive stars have typical masses of about 100 \Ms, are born in the last 100 - 200 kyr and have accretion rates of $\sim$ 0.001 \Ms\ yr$^{-1}$. The masses of all stars in our simulations are listed in Table~\ref{tab:tbl-1}.  

The maximum resolution in our simulation is 300 AU so we cannot resolve protostellar discs around individual stars. However, even if fragmentation occurs on smaller scales the clump migration timescale is shorter than the Kelvin-Helmholtz timescale at higher densities. Clumps are therefore expected to migrate inwards and merge with the central object \citep{Latif15b}. We resimulated H1 with a maximum spatial resolution of 75 AU (four times that of the others) and found that the most massive star grew to 600 \Ms\ in 57 kyr. As shown in Figure~\ref{fig:mdot}, its accretion rate and growth track are quite similar to those in the original run so our results likely hold at even higher resolution.

\begin{figure*} 
\begin{center}
\includegraphics[scale=0.8]{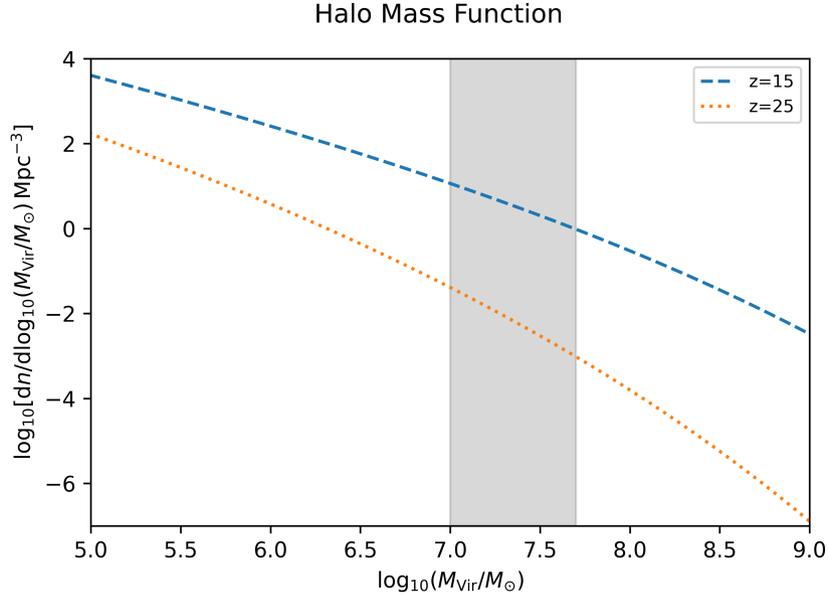}
\end{center} 
\caption{Halo mass function at different redshifts using the analytical function of \citet{War6}.}
\label{hmf}
\end{figure*}

\subsection{Comparison with previous studies}

Previous studies  investigating fragmentation in massive primordial halos under a moderate LW flux either ignored radiative feedback from  stars \citep{SS12, Latif14ApJ,Regan18b} or found it not to be important \citep[][ hereafter RD18]{Regan18b}.  RD18 only considered a single peculiar halo  at several resolutions and LW backgrounds (1, 100 and 1000 $\rm J_{21}$) for 250 kyr (and 500 kyr in two cases), a small fraction of the lifetimes of the stars in their models. Hence their results are not applicable to typical halos forming at high redshift.  The halo studied by RD18  has  a mass of $\rm \sim 10^7~\Ms\ $  and collapses at z= 24.7 while the halos in our study  collapse at $\rm13 <z<18$  except H3.  To understand this difference, we plot a halo mass function at different redshifts  in Fig. \ref{hmf} using the analytical mass function of \cite{War6}. The halo mass function shows that halos  like the halo in RD18 are very rare, 0.1 halo per 1 Mpc$^3$ while the halos in our study are a hundred times more abundant, about 10 halos per 1 Mpc$^3$. Therefore, our halos represent typical halos forming at z=15 and the RD18 halo is an outlier. Such halos  have triggered higher accretion rates in RD18 which resulted in larger stellar masses in their study.
The average mass accretion rates onto stars in our simulations are $\rm 0.001-0.01~\Ms\ /yr$ for 100 \& 500 $\rm J_{21}$ an order of magnitude lower than in RD18.  We also performed one simulation with 1000 $\rm J_{21}$ and found similar results. In fact, theoretical estimates suggest that mass accretion rates scale with sound speed $\rm \sim c_s^3/G \sim 0.1 \times (T/8000 K)^{3/2} \Ms\ yr^{-1}$.  The gas temperature in our halo centers is about 1000 K  for which the expected accretion rates are $\rm 0.004~\Ms\ yr^{-1}$,  in agreement with the results of our study. Therefore we believe our accretion rates are  realistic and in accordance with theoretical expectations.  Stellar masses in our study are about an order of magnitude lower  than those in RD18 due to the  lower mass accretion rates observed in our simulations. At such accretion rates  ($\rm < 0.04 ~\Ms\ yr^{-1}$) stars become blue and hot in our simulations and produce strong feedback which launches outflows on parsec scales and halts accretion onto them. In contrast, higher accretion rates as in RD18 result in red and cool stars which do not produce ionising radiation.  Consequently,  the mass accretion onto stars and stellar masses are an order of magnitude larger.  RD18 switched off ionising radiation in their simulations, which results in more fragmentation and three-body interactions that eject stars from the disc, terminating their growth.

We have simulated six distinct halos here that enable us to study variations from halo to halo and followed accretion onto the stars for up to 900 kyr, 4 times longer than RD18. These are thus the first simulations that have followed the accretion of the resulting objects for a relevant fraction of the lifetime of the stars in the presence of ionising radiation.  They allow us to capture the quenching of accretion onto stars because of their ionising UV flux and hence obtain more accurate final masses. We found that the  final stellar masses  are mainly determined by the stellar feedback.  Our larger ensemble of six halos therefore demonstrates that ionising UV radiation from the star sets its final mass in a wide range of environments. We have explored LW backgrounds of $\rm 100-500 ~J_{21}$ and found that Pop III stars of a few thousand solar masses can form in them. Our findings suggest that moderate LW fluxes  cannot induce the large accretion rates ($\rm \geq 0.04~\Ms\ /yr$ ) required for supermassive star formation in typical halos and massive Pop III stars of $\rm 1800-2800 \Ms\ $ form under these conditions.

\section{Discussion and Conclusion}

We find that moderate LW backgrounds in the primordial universe led to the formation of 1800-2800 \Ms\ Pop III stars, intermediate in mass to those in minihalos before such backgrounds existed \citep[30 - 500 \Ms;][]{Hirano14,hir15} and those in atomically-cooled halos in the most extreme backgrounds \citep[$\sim$ 10$^5$ \Ms; e.g.][]{tyr17}. Intermediate LW backgrounds enabled primordial halos to grow to somewhat larger masses before forming stars while allowing some H$_2$ to survive in their cores. This led to a more massive gas reservoir in the center of the halo that was at higher temperatures than those normally associated with H$_2$ cooling. At the onset of collapse, Ly$\alpha$ cooling dominated in the outer regions of the halo but H$_2$ cooling regulated the collapse of the core, but at rates that were 10 - 50 times higher than those in minihalos because the higher virial temperatures were close to the peak in the H$_2$ cooling and formation rates. Supercharged H$_2$ cooling thus produced 1000 - 3000 \Ms\ Pop III stars. We find that fragmentation in halos in moderate LW backgrounds tends to be mild so these stars are usually accompanied by a few normal Pop III stars. Our results suggest typical  stellar masses of a few thousand solar for LW strengths of 100-1000 $\rm J_{21}$. Therefore, the IMF of Pop III stars is expected to be a top heavy with the masses  up to a few thousand solar masses under such conditions. They  opened a third channel of Pop III SF that led to the birth of intermediate mass black holes in primordial galaxies.  We expect such conditions  were likely common at high redshifts as the number density of these pristine halos exposed to a given LW background  strongly varies with the strength of  the flux \citep{Agarwal14,Inayoshi15,Hartwig15,Habouzit16}. Even a factor of a few difference in LW flux changes the abundance by two orders of magnitude.

Our  simulations were evolved for more than half the expected lifetimes of the stars  \citep{Schaerer02} and radiative feedback levels off their masses well before the end of the runs. These massive stars are expected to  collapse to  BHs via the photodisintegration instability \citep{Heger02,Heger03} or He depletion in their cores without exploding \citep[see Figure 4 of][]{tyr20a}. In the future stellar evolution calculations run inline with cosmological simulations will be required to determine at what masses the stars collapse to BHs. We find that radiation from the star plays a pivotal role in its evolution in intermediate LW backgrounds. Accretion rates in halos collapsing via Ly$\alpha$ and H$_2$ cooling can be 100 times lower than those cooling by Ly$\alpha$ alone, and are close to the limit below which the stars become blue and hot rather than cool and red \citep[$\sim$ 0.02 \Ms\ yr$^{-1}$;][]{herr21a}. Thus, if a star happens to be born blue in such a halo its ionising UV radiation tends to keep it blue by curtailing accretion onto it. As the star becomes more massive it becomes more luminous and drives accretion rates even lower. None of the stars in our runs ever become red because they never reach accretion rates of 0.04 \Ms\ yr$^{-1}$. This rate is a little larger than those found in stellar evolution models to cause stars to become blue so the radiative feedback in our models should be taken to be an upper limit and the true masses of the stars may be somewhat higher.

We have assumed here  LW backgrounds  of   100 \& 500 $\rm J_{21}$ due to nearby star forming galaxies.  A stellar mass of a few times $\rm \geq10^6 \Ms$  is required to provide such flux for a Salpeter IMF with mass range of 10-100 $\rm \Ms$ \citep{Agarwal12,Dijksta14,Habouzit16,Chon17}. We assume  here $\rm T_{rad}=10^5$ K  and ignore photo-detachment of $\rm H^-$ but the spectral temperatures of the first galaxies are expected to be between $\rm T_{rad} =10^4-10^5$ K  \citep{Sugimura14,Latif15a,Agarwal15}. The LW source  halos must be located at  $\sim$10 kpc  to avoid metal pollution \citep{Dijksta14,Habouzit16} and about 60 \% of such halos are metal free at z = 15 \citep{Latif16d}. We have assumed here that the LW background is constant for the duration of our runs (about 200 Myr), and is provided by Pop II stars due to ongoing star formation in nearby halos. The impact of UV ionising radiation  from LW source galaxies was investigated by \cite{Chon17}, they found that most of it is absorbed by the filaments and dense clumps surrounding the source galaxy. In some cases, ionising radiation from the source galaxy actually promotes the collapse of the atomically-cooling halo \citep[see also][]{jet14}. The LW sources may also emit X-rays produced by the X-ray binaries or even massive stars.  X-rays heat gas at low densities but also enhance ionisation fractions that catalyze H$_2$ formation. These effects have been investigated by \citet{Jeon12}, \citet{Inayoshi12}, and \citet{Latif15a}, who found that X-rays are attenuated at higher densities and do not impact the characteristic masses of stars \citep{Hummel15}. Therefore, we expect that  ionising UV and X-rays from nearby sources  will not  have much impact on our findings.

Our accretion recipe is based on the mass influx through the accretion sphere and we assumed an accretion radius of 4 cells. Since the Jeans length must be resolved by at least four  cells in order to avoid spurious fragmentation \citep{TrueLuv}, the  accretion radius  of sink particles must not be smaller than two grid cells, see also \cite{Federrath10}. We performed an additional simulation (shown in Fig. \ref{fig:mdot})  with an accretion radius that was four times smaller and found similar stellar masses. Therefore, we do not believe that the choice of the accretion radius  has a strong impact on our results.  However, current computational constraints do not allow us to resolve flows all the way down to the surface of the star in cosmological simulations so our stellar masses should be taken to be upper limits. Instead of using an instantaneous accretion rate, we use accretion rates averaged over 1 kyr intervals and never cap the rates, but they never exceed 0.03 $\rm M_{\odot} yr^{-1}$ in our runs.

We selected six distinct halos and turned on LW backgrounds of strengths 100 $\rm J_{21}$ for H1, H3 \& H5 and 500 $\rm J_{21}$  for H2, H4 \& H6. This allowed us to robustly estimate the upper limit of stellar mass irrespective of  their merger history and environments. Our results confirm that  the upper limit of stellar mass is determined by its feedback irrespective of the strength of the LW flux and the halo mass range explored here. Therefore, if we turn on LW backgrounds of different strengths for the same halo, the results are expected to be similar. Less fragmentation occurs in our models than in simulations of normal Pop III star formation \citep{turk09,Clark11,Greif12,Susa14,Latif15a,Stacy16,Hosokawa16,Susa19,Sugimura20}. Even though SF in our halos is also regulated by H$_2$ cooling, the gas at the center of the disc is several times hotter so it is better supported by thermal pressure against fragmentation. In reality, most simulations of atomic collapse performed to date probably overestimate fragmentation because they ignore magnetic fields that likely arose in most primordial halos because of subgrid turbulent dynamos \citep{Schobera,Turk2012,Latif2013a,LatifMag2014,Sharda20}. Such fields would tend to stabilize the disc and suppress fragmentation.

It is not clear if the extremely massive Pop III stars in our simulations could later evolve into the first quasars because \citet{Smidt17} found that they must be seeded by BHs of at least 10$^5$ \Ms\ at $z \sim$ 20 to reach 10$^9$ \Ms\ by $z \sim$ 7 in the cold accretion flows that are thought to fuel their growth \citep[see also][]{Latif20b}. If not, they could instead yield a population of less-massive, lower-luminosity quasars that are yet to be discovered. Synergies between {\em JWST} and {\em Euclid} or the {\em RST} could reveal the existence of these objects when they inaugurate the era of $5 < z < 15$ quasar astronomy in the coming decade.

\section*{Acknowledgements}

MAL thanks the UAEU for funding via UPAR grant No. 31S390. DRGS thanks for funding via the Chilean BASAL Centro de Excelencia en Astrofisica y Tecnologias Afines (CATA) grant PFB-06/2007. DRGS acknowledges financial support from Millenium Nucleus NCN19\_058 (TITANs).




\section{Data Availability Statement}
The data underlying this article will be shared on reasonable request to the corresponding author.

\bibliography{smbhs.bib}





\bsp	
\label{lastpage}
\end{document}